# Spatial filtering of light by chirped photonic crystals

**Kestutis Staliunas**[1] **and Víctor J. Sánchez-Morcillo**[2]

[1]ICREA, Departament de Fisica i Enginyeria Nuclear, Universitat Politècnica de Catalunya, Colom, 11, E-08222 Terrassa, Barcelona, Spain

[2]Instituto de Investigación para la Gestión Integrada de Zonas Costeras, Universidad Politécnica de Valencia, Ctra Natzaret-Oliva S/N 46730 Grao de Gandía, Spain

*Abstract:* We propose an efficient method for spatial filtering of light beams by propagating them through 2D (also 3D) longitudinally chirped photonic crystals, i.e. through the photonic structures with fixed transverse lattice period and with the longitudinal lattice period varying along the direction of the beam propagation. We prove the proposed idea by numerically solving the paraxial propagation equation in refraction index-modulated media, and we evaluate the efficiency of the process by plane-wave-expansion analysis. The technique can be applied to filter (to clean) the packages of atomic waves (Bose condensates), as well improve the directionality of acoustic and mechanical waves.

PACS numbers: 42.55.Tv, 42.25.Fx, 42.79.-e

Spatial filtering is broadly used to improve the spatial quality of light beams [1]. The conventional technique uses a lens to form the far field image of the beam in the focal plane, and a diaphragm of the appropriate diameter to block the undesired high frequency components of the spatial spectrum, corresponding to the "noisy" part of the beam [2]. In this letter another, alternative method for spatial filtering is proposed. The new method is based on the propagation of the beam through a specially designed photonic crystal (PC), where the undesired spatial spectra components are strongly deflected, whereas the "clean" part of the beam (corresponding to the central part of the spatial spectrum) propagates through (and behind-) the crystal without deflection.

We describe the method of spatial filtering in PCs along the letter, and demonstrate its feasibility by numerical integration of the paraxial model of the wave propagation accounting explicitly for the spatial modulation of the refractive index. We also evaluate the efficiency of the process and the parameter range by an analytical-numerical study in the framework of a truncated plane-wave expansion.

The proposed method is extendable to other types of waves in the nature. As the effect of spatial filtering is indirectly related with the effect of self-collimation or subdiffraction [3-5], then the effect of filtering could be expected in all systems where the subdiffraction is present. Such systems are e.g. the Bose condensates (BECs) in spatio-temporally periodic potentials, where matter wave packets can behave subdiffractively [6,7] (here the propagation along the longitudinal direction in optics corresponds to the evolution in time for BECs), or the periodic acoustic systems, where the mechanical (sound) waves can propagate without diffraction [8]. In this way the predicted phenomenon is common for waves of different nature propagating in periodically modulated materials, and thus bears a universal character.

The propagation of light in a material with refraction index modulation is described in the paraxial approximation:

$$\frac{\partial A}{\partial Z} = i\left(\frac{\partial^2}{\partial X^2} + 4f\cos(Q_\| Z)\cos(X)\right)A \tag{1}$$

where $A(X,Z)$ is the slowly-varying envelope of the electromagnetic field. The normalisations are adapted from [5], where $X = xq_\perp$ is the normalized transverse coordinate ($q_\perp$ is the transverse component of the wavevector of index modulation); $Z = zq_\perp^2/2k_0$ is the normalized longitudinal coordinate ($k_0 = 2\pi/\lambda$ is the wavenumber of the incident monochromatic light); $Q_\|(Z) = 2q_\|(z)k_0/q_\perp^2$ is the normalized longitudinal component of the wavevector of the index modulation (longitudinal PC lattice constant), which is considered to depend on $Z$. In this way the (nearly) periodic structure of the PC contains a longitudinal chirp. The normalized modulation parameter $f = \Delta n\, k_0^2/2q_\perp^2$ is related to the refraction index modulation in original variables via $n(x,z) = n_0[1 + \Delta n\, \cos(q_\perp x)\cos(q_\| z)]$. The modulation is considered harmonic without losing the generality. The relation of Eq. (1) with the physical parameters in other systems can be found e.g. for Bose condensates in [6], and for acoustic waves in [8].

For the analytical study we assume that the chirp is weak, $|dQ_\parallel/dZ| \ll 1$ (or equivalently $|dq_\parallel/dz| \ll 4k_0^2/q_\perp^4$ in the original variables), and expand the field in harmonics of the PC structure (analogously to the unchirped case e.g. in [5]):

$$A(X,Z) = e^{iK_\perp X} \sum_{l,m} A_{l,m}(Z) e^{i\vec{Q}_{l,m}\vec{R}} . \qquad (2)$$

where $\vec{Q}_{l,m} = (l, mQ_\parallel)$, $l,m = ..-2,-1,0,1,2...$ is the set of the lattice vectors, $\vec{R} = (X,Z)$ in normalized space. The expansion (2) converts Eq. (1) into a coupled system for the evolution of the amplitudes of the harmonics,

$$\frac{dA_{l,m}(Z)}{dZ} = -i\left((l+K_\perp)^2 + mQ_\parallel(Z)\right) A_{l,m}(Z) + i f \sum_{|r-l|=1,|p-m|=1} A_{r,p}(Z) \qquad (3)$$

where the Z-dependence of $Q_\parallel(Z)$ mimics the longitudinal chirp. In the absence of chirp $A_{l,m}(Z) \propto e^{iK_\parallel Z}$, and the solvability of Eq.(3) results in the transverse dispersion relation of the PC, $K_\parallel(K_\perp)$.

For the study of self-collimation, as shown in [5,6] the expansion (2) can be truncated to the three most relevant harmonics (five harmonics in the 3D case [7]), those the closest to the mutual resonance, and the far-from resonance harmonics can be neglected. We adopt this approximation also here by considering the central harmonics $\vec{Q}_{0,0} = (0,0)$ and the first sideband components $\vec{Q}_{-1,\pm 1} = (-1,\pm Q_\parallel)$, corresponding to plane waves with propagation wavevectors $\mathbf{K} = (K_\perp, K_\parallel)$ and $\mathbf{K} = (K_\perp \pm 1, K_\parallel + Q_\parallel)$, respectively, as their dispersion curves cross at the point $\mathbf{K} = (0,0)$ for $Q_\parallel = 1$ (the triple-cross point).

The dispersion curves of the harmonic components and of the Bloch modes (the coupled states of the harmonic components) are shown in Fig.1 by the dashed and solid lines. The Bloch mode most relevant to our analysis displays, at particular frequencies, a plateau in the central part, around $K_\perp \approx 0$. This property is at the root of the self-collimation effect [3-8], as the plane wave components the beam with wavevectors lying on the plateau do not dephase in the propagation and do not lead to the diffractive spreading of the constituent beam. Most significant to our study is that the same dispersion curve also displays strongly pronounced slopes. The basic idea of the letter is that these slopes could be used for spatial filtering, as the radiation lying on the slopes is strongly deflected in the propagation through the PC. The propagation is directed along the gradient of frequency: $\mathbf{v} = d\omega/d\mathbf{K}$, therefore is perpendicular to the dispersion curve in the space $\mathbf{K} = (K_\perp, K_\parallel)$, as illustrated by thick arrows in Fig.1.

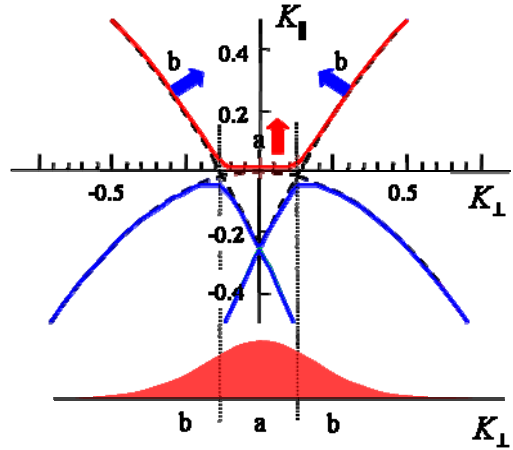

*Fig.1.* The dispersion curves of the Bloch modes in (unchirped) PC (solid lines) and of the plane wave components (dashed lines). At the bottom, the spatial spectrum of the beam consisting of (a) the central (regular) part, and (b) the wings (the part to be filtered) is illustrated. The gray areas around the dispersion curves of the Bloch modes illustrate the density of the occupation of the corresponding modes. Parameters: $Q_\parallel = 0.75$, $f = 0.04$

The problem is, however, that the incoming radiation projects very weakly into the Bloch modes in the region of the slopes, because in these segments the plane wave components are almost orthogonal to the corresponding Bloch modes. The efficiency of the projection into the Bloch modes is inversely proportional to the distance between the dispersion curves of the Bloch modes (solid lines), and the dispersion curves of corresponding plane waves (dashed lines) [9]. The grey areas on the dispersion curves of the Bloch modes in Fig.1 illustrate the efficiency of the projection.

The projection problem can be overcome using the longitudinally chirped PC. The idea is illustrated in the Fig.2. The longitudinal period at the entrance is chosen as corresponding to Fig. 2(a), in order to project efficiently all the radiation from the entire spatial spectrum of the beam onto the basic Bloch mode. Along the PC the period decreases ($Q_\parallel(Z)$ increases) and the dispersion diagram deforms continuously (illustrated by the arrows in Fig.2). At the rear face of the PC the period is such that the dispersion curve attains the shape shown in Fig. 2(b). If the chirp is sufficiently weak, the process can be considered adiabatic. Then, the occupations of the Bloch modes remain unchanged and all the excitation is transferred to the slopes, and eventually deflected.

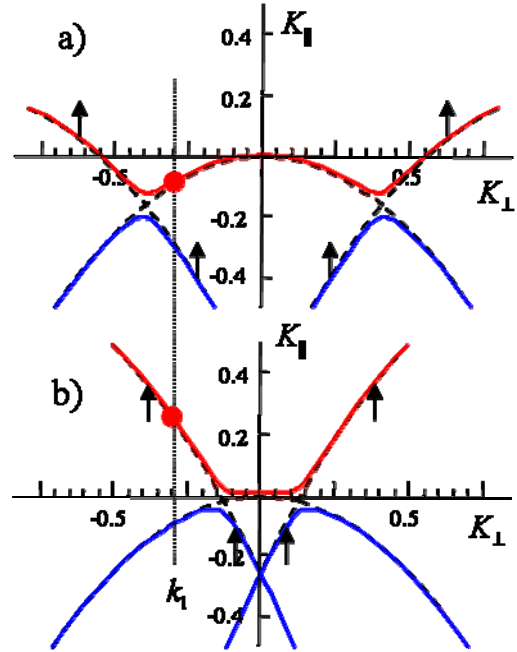

*Fig.2. The dispersion curves in the PC with chirp: (a) at the entrance, where $Q_\parallel(0) = 0.2$, (b) at the rear face, where $Q_\parallel(L) = 0.75$. The gray areas around the dispersion curves of the Bloch modes illustrate the density of the occupation of the corresponding modes. $f = 0.04$*

We prove the above described idea by the numerical simulation of the paraxial model Eq. (1) in 2D demonstrating the predicted effect of spatial filtering. The results are presented in Fig.3. In Figs. 3(a)-(b) the input is a narrow beam, with sufficiently broad spatial spectrum (broader than the width of plateau at the rear face of the PC). When the PC is without chirp, the filtering is absent, and only narrow areas from the central maximum of spatial spectrum are deflected [Fig. 3(a)]. The dips occur at the cross points of the dispersion curves of the harmonic components and are due to two reasons:

the projection of the plane waves into the Bloch modes at the entrance of the PC (the plane waves project equally on both Bloch modes at the cross point of the dispersion curves of harmonics), and the back-projection of the radiation of the Bloch modes into the plane waves at the rear face of the PC.

In case of the chirped PC (Fig.3.b) the radiation projects efficiently onto the slopes, as described above, and during the propagation through the PC is transported to the sidebands and deflected with almost 100% efficiency.

As the effect is linear then the different angular components of the radiation can be treated separately, and the process, illustrated above with the regular beams, works equally well with the noisy beams. Fig. 3(c) shows the propagation of a beam with spatial phase noise, with approximately the same spatial spectrum width as in Figs. 3(a)-(b). As expected, the filtering works, resulting in a cleaning of the beam.

The ideal filtering can be realized in the adiabatic limit, when the chirp is very weak, and respectively the PC is extremely long. In practice, however, the limit is difficult to realize, and therefore it is significant to evaluate the minimum length of the PC that can ensure the reasonable filtering. The process of projection of the radiation onto the slopes is described by Eqs. (3). By fixing a particular angle $K_\perp$ (indicated in Fig.2 by a vertical dotted line) one can restrict the analysis to that of two coupled oscillators (two harmonics) participating in the excitation exchange process. The other harmonic, being far from resonance does not interfere the process under the study. From Eqs.(3) we obtain

$$\frac{dA_1(Z)}{dZ} = ifA_2(Z) \tag{4.a}$$

$$\frac{dA_2(Z)}{dZ} = i\Delta Q_\parallel(Z)A_2(Z) + ifA_1(Z) \tag{4.b}$$

where $(A_1, A_2) = (A_{0,0}, A_{-1,-1})e^{ik_\perp^2 z}$. The reference propagation wavenumber is fixed to that of the zero harmonics, and $\Delta Q_\parallel(Z) = Q_\parallel(Z) - (K_\perp - 1)^2 + K_\perp^2$ is the mismatch between the propagation wavenumbers of both harmonics.

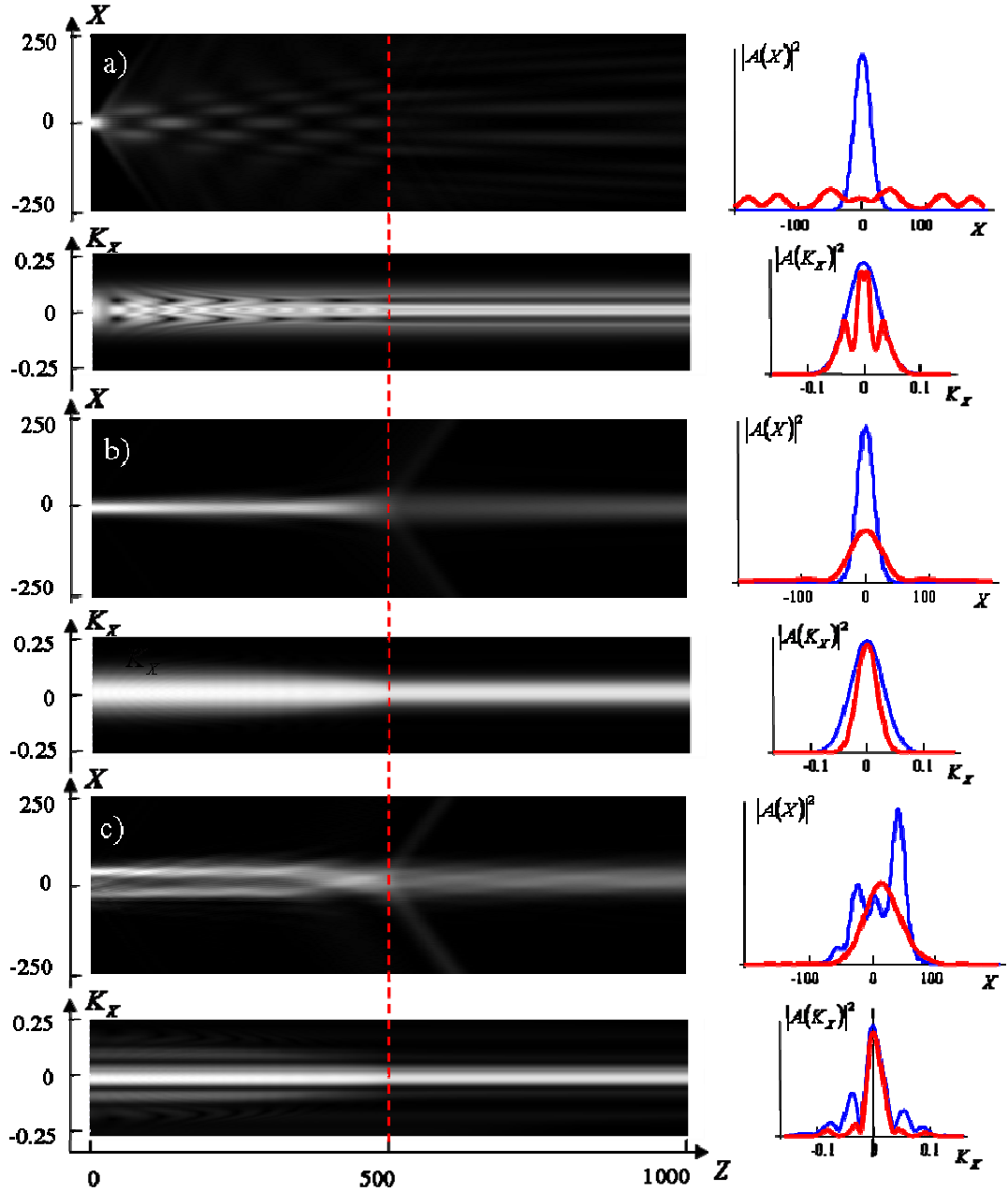

*Fig.3.* Propagation of the beams through PCs. (a) Regular beam propagating through the unchirped PC with $Q_\parallel = 0.92$, and (b) through the chirped PC with $Q_\parallel(0) = 0.61$, $Q_\parallel(L) = 0.92$; (c) noisy beam, with the width of the spatial spectrum approximately equal to that in cases (a,b) propagating through the chirped crystal as in (b). In all cases $f = 0.022$. In the insets field distributions are shown (at the entrance of the crystal at Z=0 (thin, blue line) and well behind the crystal at Z=1000 (thick, red line)) both in space- and spatial spectra domain. The dashed line indicates the rear face of the crystal (at Z=500). The normalized chirp parameter is $\beta = \alpha/f^2 = 1.24$ in cases (b,c).

The scenario of energy exchange is as follows: initially the first oscillator is excited (i.e. the one belonging to the harmonics $\vec{Q}_{0,0} = (0,0)$), then the frequency of the second oscillator (belonging to $\vec{Q}_{-1,\pm 1} = (-1, \pm Q_\parallel)$) is varied along the propagation, as illustrated in Fig.4.a. We expect that at the end of the procedure the radiation transfers to the second oscillator, equivalently, the excitation remains on the same Bloch mode. Fig.4.b shows the evolution of the excitation of the oscillators for a linear chirp $\Delta Q_\parallel(Z) = \alpha Z$. In the weak chirp case ($\alpha \ll f^2$) the excitation indeed remains on the same Bloch mode (equivalently, transfers from one oscillator to the other). The transfer occurs essentially in the (longitudinal) segment of the PC corresponding to the resonance between the interacting harmonics, i.e. when $|\Delta Q_\parallel(Z)| \leq f$. In the case of the strong chirp the situation is opposite: the radiation is exchanged between the both Bloch modes, and equivalently remains on the same harmonics. The latter situation is related with the Landau-Zener-like tunnelling during the fast (nonadiabatic) processes [10].

The parameters of Eqs. (4) can be scaled, leading to only one free parameter, the normalised strength of the linear chirp $\beta = \alpha / f^2$. This implies that the excitation transfer (and eventually the efficiency of filtering) depends solely on that parameter. A numerical study of the efficiency of the excitation transfer, based on Eqs. (4), shows that the e.g. a projection of 50% of the excitation to the slopes occurs at $\beta_{50\%} \approx 9 \pm 0.2$, a projection of 90% occurs at $\beta_{90\%} \approx 2.7 \pm 0.05$, and a projection of 99% occurs at $\beta_{99\%} \approx 1.25 \pm 0.02$ [11].

We estimate the physical parameters for the filtering. In order to make the effect of filtering useful (to filter out the substantial part of the spatial spectrum of the beam) the condition $|\Delta Q_\parallel| \approx 1$ is to be fulfilled, which means that the cross-point of dispersion curves moves over nearly all the spatial spectra. Assuming the limit of weak modulation $f \ll 1$, this leads to the estimation for the PC length (in terms of longitudinal periods of the PC $N$): $N \geq (2\pi \beta f^2)^{-1}$, or in terms of original variables: $N \geq 2(k_0/q_\perp)^4 / (\pi \beta \Delta n^2)$. We estimated the PC filter thickness for several cases: i) for the usual glasses with weakly modulated index ($\Delta n \approx 10^{-4}$), and $(k_0/q_\perp) \approx 10$, the 90% filtering requires $N \geq 2000$ longitudinal periods. ii) for special glasses with moderately modulated index (see e.g. [12]), or photorefractive crystals [13] ($\Delta n \approx 10^{-3}$), the 90% filtering should be observable for the PCs as long as containing $N \geq 20$ longitudinal periods. iii) for high contrast PCs ($\Delta n \approx 0.1 - 1$) the above estimation states that the filtering is efficient for an arbitrary length of the crystal, i.e. does not impose any minimum length of the PC.

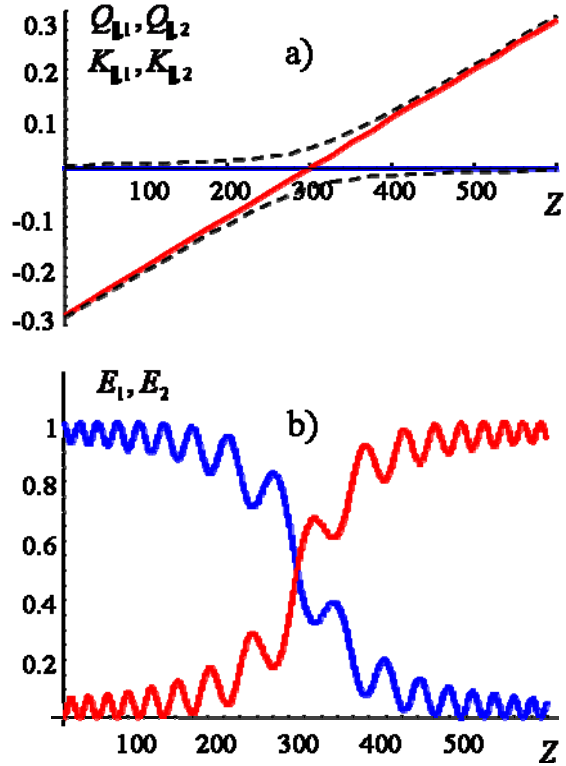

*Fig.4. (a) Variation of the longitudinal propagation wavenumbers of the harmonic components (solid lines) and of the Bloch modes (dashed lines) along the chirped PC. (b) variation of the excitation of the two harmonic components. Parameters: $\alpha = 0.001$, $f = 0.04$ (the normalized chirp is $\beta = \alpha/f^2 = 0.625$)*

In conclusion, we predict and show the possibility of spatial filtering in longitudinally chirped PCs. We evaluated the parameters of the PC, concluding that they are reasonable for the experimental demonstration, as well as for the implementation of the effect. The filtering has been demonstrated numerically in the case of harmonic modulation of the refraction index, but the idea is also applicable also to "usual" PCs, where due to the microfabrication techniques the index profile is sharp.

The presented analysis concerns the 2D case, i.e. beams with one transverse dimension propagating in 2D PC, but could be also extended into the 3D case. The mechanism of the projection on the slopes of the dispersion surfaces and the estimated values of the chirp remains the same as in 2D case. The essential difference with the 2D case is the symmetry of the lattice in the transverse plane of the [7]. It comes out from a preliminary study that the different symmetries of the PC in transverse plane result in the different shape of the spot in the spatial Fourier domain, e.g. a square lattice results in a square filter pass window. Reasonable filters for applications should be of hexagonal or octagonal symmetry.

The method can be optimized. In order to achieve a larger projection into the harmonics not only the longitudinal period, but also the other parameters could be varied. E.g., as shown in [14], the slow variation of the index modulation amplitude (the hole radius) can improve the transfer of energy. Also a special function of the chirp can be used, instead of the linear one considered in the present letter.

We note, that the effect of filtering is applicable for other sorts of waves. The filtering of sound waves follows straightforwardly from the above study due to similarity of waves in these systems. The cleaning of the BECs requires, however, a special geometry. As the propagation in optics corresponds to the time evolution in BECs, then the 2D PC corresponds to the modulated in time and space potential of the BEC, as discussed in [6,7]. The longitudinal chirp in optics then corresponds to the smooth (adiabatic) variation of the temporal modulation frequency in BECs.

The work was financially supported by Spanish Ministerio de Educación y Ciencia and European Union FEDER through projects FIS2005-07931-C03-02, and -03